\pdfoutput=1

\documentclass[aps,prx,twocolumn,superscriptaddress]{revtex4-1}

\usepackage{caption}
\usepackage{subcaption}
\usepackage{mathrsfs}
\usepackage{nicefrac}
\usepackage{amssymb}
\usepackage{comment}

\usepackage{graphicx,bm,amsmath}
\usepackage[usenames,dvipsnames]{color}
\usepackage[colorlinks,bookmarks=false,citecolor=NavyBlue,linkcolor=Red,urlcolor=blue]{hyperref}
\usepackage[bbgreekl]{mathbbol}
\usepackage{xcolor}

\usepackage{hyperref}

\usepackage{tikz}

\usetikzlibrary{arrows}
\usetikzlibrary{math}
\usetikzlibrary{calc}
\usetikzlibrary{shapes.geometric}

\usepackage{bbold}

\DeclareMathOperator{\tr}{tr}

\newcommand{\be}{\begin{equation}}
\newcommand{\ee}{\end{equation}}
\newcommand{\bea}{\begin{eqnarray}}
\newcommand{\eea}{\end{eqnarray}}

\definecolor{smoothred}{HTML}{C5232F}
\definecolor{mygreen}{rgb}{0,0.5,0}
\definecolor{myblue}{rgb}{0,0,0.75}
\definecolor{mymagenta}{cmyk}{0,1,0,0.12}

\newcommand{\english}[1]{{\color{black} #1}}

\def\doi{http://dx.doi.org/}

\begin{document}

\title{Optimizing Radiotherapy Plans for Cancer Treatment with Tensor Networks}

\author{Samuele Cavinato}
\affiliation{Dipartimento di Fisica e Astronomia ``G. Galilei'', Universit\`a di Padova, I-35131 Padova, Italy}

\author{Timo Felser}
\affiliation{Dipartimento di Fisica e Astronomia ``G. Galilei'', Universit\`a di Padova, I-35131 Padova, Italy}
\affiliation{INFN, Sezione di Padova, I-35131 Padova, Italy.} 
\affiliation{Theoretische Physik, Universit\"at des Saarlandes, D-66123 Saarbr\"ucken, Germany.}

\author{Marco Fusella}
\affiliation{IOV-IRCCS, I-35128 Padova, Italy}

\author{Marta Paiusco}
\affiliation{IOV-IRCCS, I-35128 Padova, Italy}

\author{Simone Montangero}
\affiliation{Dipartimento di Fisica e Astronomia ``G. Galilei'', Universit\`a di Padova, I-35131 Padova, Italy}   
\affiliation{INFN, Sezione di Padova, I-35131 Padova, Italy.}

\date{\today}

\begin{abstract}
We present a novel application of Tensor Network methods in cancer treatment as a potential tool to solve the dose optimization problem in radiotherapy. In particular, the Intensity-Modulated Radiation Therapy (IMRT) technique -- that allows treating irregular and inhomogeneous tumors while reducing the radiation toxicity on healthy organs -- is based on the optimization of the radiation beamlets intensities. The optimization aims to maximize the delivery of the therapy dose to cancer while avoiding the organs at risk to prevent their damage by the radiation. Here, we map the dose optimization problem into the search of the ground state of an Ising-like Hamiltonian, describing a system of long-range interacting qubits. Finally, we apply a Tree Tensor Network algorithm to find 
 the ground-state of the Hamiltonian. In particular, we present an anatomical scenario exemplifying a prostate cancer treatment. A similar approach can be applied to future hybrid classical-quantum algorithms, paving the way for the use of quantum technologies in future medical treatments.
\end{abstract}

\maketitle
\section{Introduction}
\begin{center} 
\begin{figure*}[t!]
\includegraphics[width=1.0\textwidth]{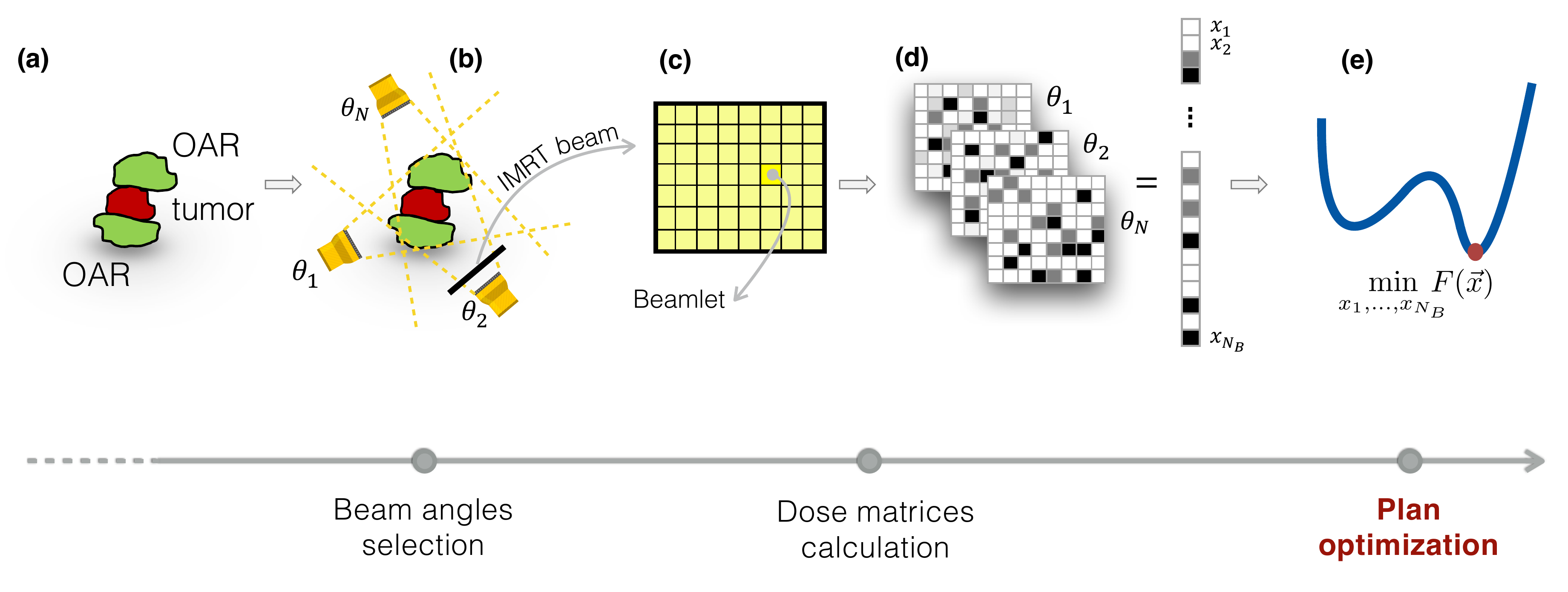}
\captionsetup{justification=centerlast}
\caption{ \label{fig:Concept}
Typical radiation therapy treatment procedure for the IMRT. The tumor (red) lies close to vital organs at risk (OARs, green) (a). In IMRT treatments the tumor is irradiated from different angles ($\theta_1, \hdots, \theta_N$) using photon beams (b). Each beam is subdivided into a grid of smaller pencil beams called beamlets in order to modulate the local beam fluence (c). A weight $x_j \geq 0$ is associated to the j-th beamlet in order to quantify its contribution to the whole beam (d). The $x_j$s becomes the optimization variables of a cost function $F(x_1, \hdots, x_{N_{B}})$ which expresses the distance between the desired dose and the delivered dose and the final aim of the RT planning procedure is to find the configuration $\vec{x}$ of the beamlets with minimizes the cost function (e).}
\end{figure*}
\end{center}
Radiotherapy is one of the techniques used to treat solid tumors by means of a ionizing radiation. The radiation dose released into the cancer tissue damages the DNA of the tumor cells leading to their death or slowing down the growth of the tumor~\citep{biological-damage-RT, biological-effect-RT-cancer}. Radiotherapy is often used in combination with other therapies like surgery, chemotherapy or immunotherapy to improve their global efficacy~\cite{radio+chemo, radio+immuno, RT+chemio, RT+immunotherapy}.

One of the hardest challenges encountered while treating patients with ionizing radiations is to deliver an optimal dose to the targeted tumor while keeping the radiation as low as possible in the surrounding healthy tissues. One of the most frequently used techniques nowadays is the \textit{Intensity-Modulated Radiation Therapy} (IMRT) \cite{IMRT-what-is, Bortfeld2006, Taylor2018, Elith2020}. In IMRT, the applied radiation is modulated to reach an optimal dose distribution inside the patient. This optimal modulation is obtained by solving a highly non-trivial numerical \english{optimization} problem with a high number of \english{optimization} parameters and numerous constraints on the final radiation dose distribution. 
Over the past years, several numerical techniques have been developed to address this challenge~\cite{Cotrutz_2001, webbSA, Bortfeld2006, CENSOR2012109, Ezzell-1996}. In 2015, Nazareth and Spaans proposed to solve the IMRT beam fluence optimization problem using the D-Wave annealer~\cite{dwave-quantum-annealing-IMRT}, and more recently El Naqa \textit{et al.} proposed an approach exploiting the simulated quantum tunnelling effect~\cite{quantum-inspired-tunneling-IMRT}.

Despite these remarkable novel approaches paved the way for future applications of quantum computation in medicine, their application is still limited. On the one hand, this is due to the lack of scalable quantum hardware. On the other hand, it is not straightforward to extend what is done on classical computers to quantum ones
due to the lack of a robust and clear strategy to map the classical IMRT problem to quantum hardware. To try overcome these limitations as well as to further investigate the applicability of quantum-inspired techniques to the solution of classical optimization problems and foster future applications of quantum technologies to medicine, here we apply 
 Tensor Networks (TNs) to an IMRT dose optimization problem. 

TNs are one of the most successful algorithms for simulating quantum many-body systems on classical computers. Indeed, whenever possible, they efficiently represent quantum many-body wavefunctions 
in a compact form on classical computers~\cite{Schollw_ck_2011,TTNA19,SimoneBook,MERA_bible,Or_s_2014}. Il the last few decades, TNs have proven 
their effectiveness 
in the research and analysis of quantum many-body systems, especially for low-dimensional ground-state~\cite{aTTN, m07, sv13, dm16, gqh17, bauls2019simulating, felser2019twodimensional, Carmen_Ba_uls_2020}. In addition to that, thanks to the properties they share with quantum hardware, TNs may play the role of test benches for the development of quantum algorithms~\cite{Huggins_2019,kim2017robust,zhou2020limits}.

Hereafter, we show how to solve an IMRT optimization problem with TNs. We first introduce how the classical cost function is mapped into an Ising-like Hamiltonian, where the \english{optimization} variables are represented as a set of long-range interacting spins. Finally, we solve the classical \english{optimization} problem by finding the ground-state for this Hamiltonian using TNs.
We present the application of TNs to two different toy models and to a more realistic anatomical scenario simulating a prostate cancer treatment. We show that TNs results are compatible with other classical techniques, Quadratic Programming (QP) and Simulated Annealing (SA).
Our results pave the way to the application of TNs methods to 
more complex and realistic clinical scenarios, contributing to 
building solid foundations for future applications of quantum computation to medicine.
In the midterm, we foresee the development and application of  classical TN methods to the solution of the IMRT problem.  

The manuscript is structured as follows: In Section \ref{sec:Radio} we provide a brief introduction to the dose \english{optimization} problem in IMRT together with its basic mathematical description. In Section \ref{sec:TTN} the mapping procedure of the classical cost function to the Ising-type Hamiltonian is described in more detail. Furthermore, this section includes a brief introduction to TNs as well. Finally, we present and discuss the main results obtained from this study in Section \ref{sec:results}.

\section{Radiotherapy optimization problem}\label{sec:Radio}
In radiotherapy, cancer cells are treated by releasing a certain amount of radioactive dose inside the tumor. 
Modern radiotherapy offers various techniques for treating tumors~\cite{SBRT-what-is, VMAT-what-is, image-guided-RT, brachitherapy-what-is, advances-RT-general} and the choice of one of them depends on factors like the site of the disease, type of cancer and overall patient's conditions~\cite{PMID:30373115, Zhang2020, Delaney2017}. 
One of the most impacting techniques in terms of improvement of treatments quality is the IMRT which became clinical available in its first implementations in the early 2000s~\cite{Bortfeld2006,IMRT-status-2001}.

The goal of IMRT is to create a personalized dose distribution for each patient's anatomy that ensures the appropriate dose for the tumor while saving the Organs At Risk (OARs) as much as possible. The choice of IMRT treatment is nowadays mandatory in very challenging cases where a high dose is required for very irregular tumor shapes surrounded by critical OARs. In order to effectively treat this kind of diseases, the IMRT benefits from a non-uniform intensity distribution of the radiation beams. In particular, the radiation beams are modulated by dividing the fluence of each treatment beam into a certain number of smaller pencil-beams called \textit{ beamlets} that can be delivered through the movement of the Multileaf Collimator (MLC).

However, finding the ideal intensity of every single beamlet for a desired treatment which optimises the relation between fields arrangement and dose distribution inside the patient is a highly complex problem. This optimisation problem is typically solved as an \textit{inverse problem} which is encoded into a cost function to be minimized in order to find the optimal beamlet configuration.

In the past years, different approaches have been proposed, among them, the analytic transform method~\cite{Bortfeld1995}, algebraic solutions~\cite{CENSOR2012109}, gradient descent~\cite{Cotrutz_2001}, genetic algorithms~\cite{Ezzell-1996} and simulated annealing~\cite{webbSA}. Although these powerful methods have been used in everyday clinical practice with increasing success, finding patient-specific plans is still an open problem since it should account for many factors, increasing the complexity of the problem. Therefore, new mathematical and physical solutions need to be developed in order both to spare precious computational time and to improve the quality of the treatments delivered, with the aim of enhancing our capability to save human lives in the fight againts cancer.

In what follows, we present the main elements of radiotherapy, explaining the general planning procedure for an  IMRT  treatment, focusing on the underlying numerical optimization problem. 

Fig.~\ref{fig:Concept} illustrates a typical treatment procedure. The targeted tumor (red), located between two OARs (green), is irradiated by  several photon beams from different angles $\theta_k$  (see Fig. \ref{fig:Concept}b). The underlying geometry is defined in the dosimetric plans created by the medical physicist. The desired dose to the target and the OARs are defined in the treatment plan, following the goal of to spare healthy organs while the proper dose is delivered to the target.
\begin{center} 
\begin{figure*}[t]
\centering
\includegraphics[width=1.0\textwidth]{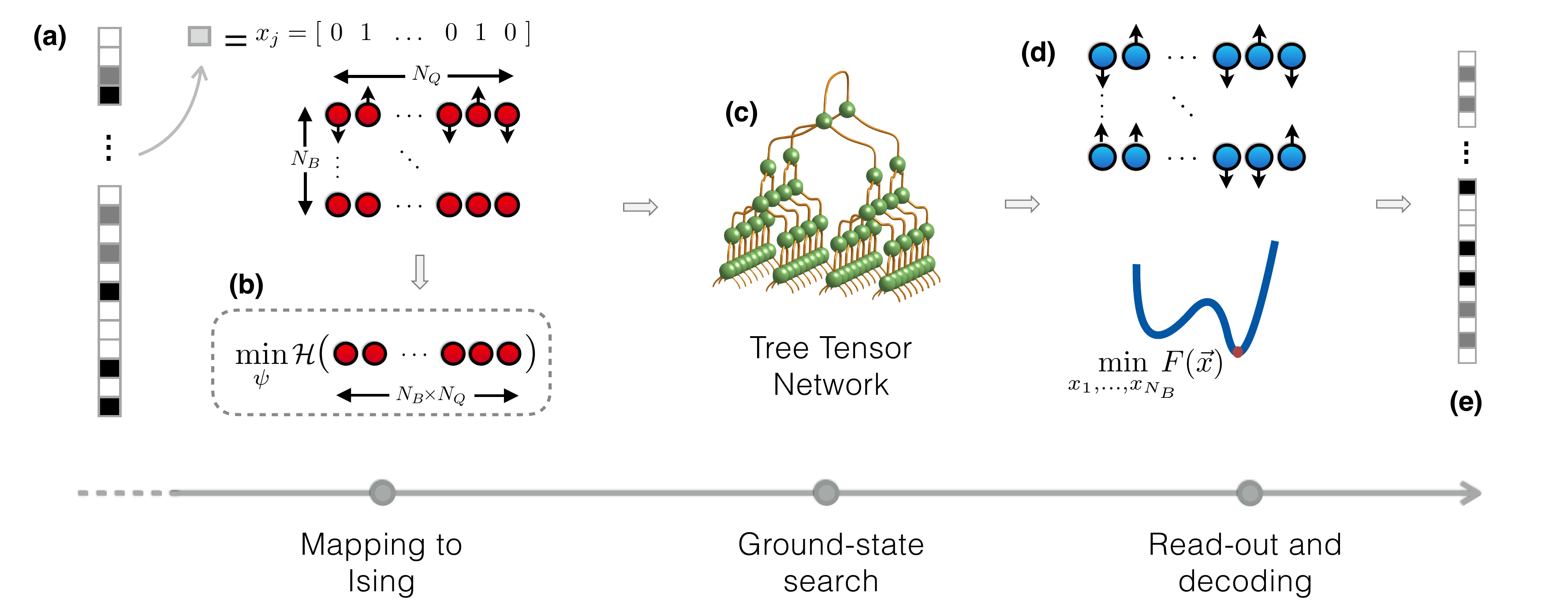}
\captionsetup{justification=centerlast}
\caption{\label{fig:mapping} Solving the classical IMRT optimization problem as a quantum Hamiltonian with TNs. The beamlet weights, $x_j$, are represented as a set of long-range pairwise interacting spins (a). The initial problem of minimizing a cost function is thus mapped into a ground-state search problem for the Ising-like Hamiltonian (b) which can be efficiently solved using the Tree Tensor Network algorithm (c). After the minimization the final spin configuration we read out the results by reconstructing the optimal values for the beamlet weights (d, e).}
\end{figure*}
\end{center}

To achieve this treatment goal, the IMRT optimization process begins with the splitting of each beam fluence into a grid of {beamlets} as shown in  Fig.~\ref{fig:Concept}c. Only beamlets that traverse the target can be optimized, and the intensity of each $j$-th beamlet can be modulated independently with a weight $x_j$. (Fig. \ref{fig:Concept}d). The volume of the patient considered for the optimization can be divided in 3D finite-size elements called \textit{voxels}. The contribution of each beamlet to each voxel depends on their intensity as well as on the geometry and physical properties of each beam and patient's characteristics. 

All these information are gathered in the \textit{influence matrix} A which maps every beamlet to each voxel. The influence matrix is usually provided by the Treatment Planning System (TPS) used in the clinical environment, and a complete description of its computation goes far beyond the aim of this work.

The sum of all beamlets produces the total dose within a given voxel $i$ and contributes to the definition of a global dose distribution $D(\bf x)$. Thus, the total dose delivered to the voxel $i$ can be expressed as: 
\be
D_i({\bf x}) \equiv D_i(x_1, \hdots ,x_{N_B}) = \sum_{j = 1}^{N_B} a_{ij}x_j ~,
\label{eq:delivered_dose}
\ee
where $N_B$ is the total number of beamlets and $a_{ij}$ describes the so-called \textit{influence matrix}, $A$, giving the unmodulated contribution of the $j$-th beamlet to the $i$-th voxel. 

The IMRT planning procedure is then solved through an iterative inverse planning process: the  intensity of  the  applied  photon beams are optimized towards the prescribed dose distribution $D^{(P)}$ inside the patient. The dose criteria are typically defined using Dose- Volume Histograms (DVH). In particular, we minimize the distance between the delivered dose $D(\bf{x})$ and the desired  dose in the  patient $D^{(P)}$ in  the discretized volume with the beamlet weights {\bf x} being the optimization variables (see in Fig. 1e).  We will describe the process through the following quadratic cost function:
\be
F({\bf x}) = \sum_{r = 0}^{R} \sum_{i = 1}^{{\cal{V}}_r} \gamma_i \left[D_i({\bf x}) - D_i^{(P)} \right]^2
\label{eq:cost_function}
\ee
where $r$ is an index running over all the volumes (the targeted tumor and the OARs), and $\gamma_i$ a weight assigned to the $i^{th}$ voxel in order to prioritize certain volumes during the treatment. ${\cal{V}}_r$ gives the total number of voxels belonging to the object $r$.

\section{Solving classical problems with Tensor Networks}\label{sec:TTN}
\subsection{Mapping the problem to a classical Hamiltonian}\label{subsec:mapping}
In the following, we describe how to rewrite the optimization problem in Eq.~(\ref{eq:cost_function}) into a ground-state search of a quantum many-body Hamiltonian. In particular, we here propose a procedure based on the binary-decimal conversion to map the cost function $F({\bf x})$ into an Ising-type Hamiltonian, ${\cal{H}}_{IMRT}$. The procedure is summarized in Fig. \ref{fig:mapping}.

We first discretize the weights $x_j$ for each beamlet by a set of $N_Q$ bits. $N_Q$ is called the \textit{bit-depth}. Thus, we represent
\be
x_j \approx \frac{1}{(B/2)} \sum_{n = 1}^{N_Q} 2^{n-1} b_{n}^{(j)}
\label{eq:decimal_binary}
\ee
with the bits $b_{n}^{(j)} = \{0,1\}$, and introducing a normalization constant $B$ to set the range of the beamlets weights such that $x_j \in \left[0;\frac{2^{N_Q}-1}{(B/2)}\right]$. With increasing number of bits $N_Q$ we can increase the resolution of the discretization. Then, we map the binary values $b_{n}^{(j)}$ into spin variables $s_{n}^{(j)} = \{-1;+1\} = 2b_{n}^{(j)} - 1$ for each site $j$. Consequently, we construct a $N_B\times N_Q$-dimensional many-body Hamiltonian with the first dimension running over the different beamlets and the second representing the discretized space for each beamlet. Inserting the spin variables together with Eq.~(\ref{eq:decimal_binary}) into the cost function of Eq.~(\ref{eq:cost_function}), we obtain this Ising-type Hamiltonian with the following general expression:
\be
{\cal{H}}_{IMRT} = {\cal{H}}_{SP} + {\cal{H}}_{INT}^{(a)} + {\cal{H}}_{INT}^{(b)}
\label{eq:ising_type}
\ee
where the first term describes the single particle terms with
\be
{\cal{H}}_{SP} = \sum_{j,n} \left[ \sum_i \frac{\gamma_i}{B}\left( a_{ij}\sum_k a_{ik} - 2D_i^{(P)}a_{ij}\right)2^{n-1}\right] s_n^{(j)} ~,
\label{eq:single_particle}
\ee
the second term captures spins interacting in the same beamlet
\be 
{\cal{H}}_{INT}^{(a)} = \sum_j\sum_{m\neq n}\left[\sum_i\gamma_i \frac{ a_{ij}^2}{B^2}2^{n-1}2^{m-1}\right] s_n^{(j)}s_m^{(j)} ~,
\label{eq:int_same_b}
\ee
and the last term represents the interactions between different beamlets
\be
{\cal{H}}_{INT}^{(b)} = \sum_{j\neq k}\sum_{n,m}\left[\sum_i\gamma_i \frac{a_{ij}a_{ik}}{B^2}2^{n-1}2^{m-1}\right] s_n^{(j)}s_m^{(k)} ~.
\label{eq:int_diff_b}
\ee

We point out, that this Hamiltonian describes a two-dimensional fully-connected lattice of long-range interacting spins with $N_B$ sites on one direction and $N_Q$ on the other. At this point, we can solve the initial problem of minimizing Eq.~(\ref{eq:cost_function}) by finding the ground-state for the classical Ising-type Hamiltonian in Eq.~(\ref{eq:ising_type}). 

The ground state of the system is given by the classical spin-configuration which provides the lowest energy $E_0$, that is the configuration that minimizes Eq. (\ref{eq:cost_function}). The case $E_0=0$ corresponds to the optimal beamlet setup which results in exactly the desired dose distribution within the patient.Anyhow, in practice, the exact desired dose distribution is not always achievable for the optimization. Thus, we can find the solution of the initial optimization problem by going through all possible spin-configurations of the Ising-type Hamiltonian.  Anyhow, this classical search rapidly becomes unfeasible, as the number of spin-configurations grows exponentially with increasing system size. And since the number of beamlets required in a radiotherapy treatment can easily get in the order of a few thousand, the number of spins required to represent them is practically too high to solve the problem efficiently in the classical regime. For this reason, in the following section we introduce an approach based on TNs to address this complex optimization task.

\subsection{Description of TTN for solving quantum many-body systems}\label{subsec:qmb}

Considering the classical Hamiltonian ${\cal{H}}_{IMRT}$, we further allow each spin to be a quantum variable by representing $s_{n}^{(j)}$ with the Pauli matrix $\sigma_z$. In this way, the quantum Ising-type Hamitonian ${\cal{H}}_{IMRT}$ is a diagonal matrix with its entries corresponding to the energies of all possible spin combinations of the classical Hamiltonian. Here, TNs are a vital tool for finding ground states and their physical properties despite the exponentially growing Hilbert space~\cite{SimoneBook,Schollw_ck_2011,NiC00}. In the following, we introduce the idea of TNs to address this challenging task of investigating complex quantum many-body systems. For a more in depth introduction on TNs we refer to sophisticated literature~\cite{SimoneBook,Schollw_ck_2011,Or_s_2014,TTNA19}

TNs are used to efficiently represent quantum many-body wavefunctions $|\psi\rangle$, which live in the tensor product $\mathcal{H} = \mathcal{H}_1\otimes \mathcal{H}_2\otimes \cdots \mathcal{H}_N$ of $N$ local Hilbert spaces $\mathcal{H}_k$, each assumed to be of finite dimension $d$. Expressing such a state in real-space product basis means decomposing the wavefunction as
\begin{equation}
|\psi \rangle = \sum_{i_1,...i_L = 1}^{d}{c_{i_1,...,i_L} |i_1\rangle_1 \otimes |i_2\rangle_2 \otimes ...\otimes |i_L\rangle_L} ~,
\label{TN-eq:totalstate}
\end{equation}
where $\{ |i\rangle_k \}_i$ is the canonical basis of site $k$, spanning $\mathcal{H}_k$.
The exact description of such a general state by all possible combinations of local states requires $d^N$ coefficients $c_{i_1,...,i_N}$. Thus, the number of coefficients increases exponentially with the system size $N$ in the exact representation of the wave-function.

TNs offer a more efficient representation by decomposing the complete rank-$N$ tensor (containing all $d^N$ coefficients) into a set of local tensors with smaller rank, connected with auxiliary indices. We control the dimension of the auxiliary indices with the bond-dimension $\chi$ and thereby the amount of captured information. Thus, tuning this parameter $\chi$, TNs interpolate between a product state, where quantum correlations are neglected, and the exact, but inefficient representation.

The decomposition of the complete rank-$N$ tensor can be executed in several ways giving rise to different Tensor Network geometries. The most prominent TN representations are the Matrix Product States (MPS) for 1D systems \cite{_stlund_1995, Schollw_ck_2011, MPSfaithful}, which addresses all sites with one corresponding tensor and their two-dimensional variant, the Projected Entangled Pair States (PEPS) \cite{PEPSfirsttime2004, PEPSbasics2006, Or_s_2014}. Tree Tensor Networks (TTN)~\cite{VidalDuanTTN2006,HomogeneousTTN,TTN14,gqh17} with their hierarchical structure can in principle be defined in any lattice dimension. As an illustration of the representation power, the MPS, for instance, reduces the number of parameters to the upper bound of $Nd\chi^2$ controlled by the chosen bond-dimension $\chi$, leading to a linear dependence on the system size $N$ rather than an exponential one.

 In our analysis, we use the latter, a TTN. The TTN offers better connectivity between long-range interactions (with a logarithmic distance threw the network), while for the simpler MPS the distance by connecting tensors within the network is linear. In contrast to the PEPS, the TTN can be optimized with a lower computational complexity as well ($\mathcal{O}\left( \chi^4 \right)$ for the TTN vs. $\mathcal{O}\left(\chi^{10}\right)$ for the PEPS)
 
Due to the bond-dimension $\chi>1$, we perform a quantum ground state search within the subspace limited by $\chi$. Thus, in contrast to the classical optimization routines, we explore within one optimization step several classical solutions as they are superposed in the quantum representation of the TTN. This allows us further to tunnel through higher, but reasonably thin, potentials within the optimization landscape.

After converging to the quantum ground state, we in general still obtain a superposition of classical solutions. As we know, that all solutions to the problem are classical, in theory, when the TTN algorithm is fully converged to the ground state, each of the superposed classical solutions separately has the same ground state energy $E_0$. Due to this possible degeneracy, we can select one classical solution from the TTN in the following way: We truncate the bond-dimension down to $\chi=1$, leaving us with a separable, mean-field solution
\[
|\Psi_{\chi=1}\rangle =  |\psi_1\rangle \otimes |\psi_2\rangle \otimes ...\otimes |\psi_L\rangle ~,
\]

in which the only superposition can be local (such as a local site being $|\psi_n\rangle = 1/\sqrt{2}(|\uparrow \rangle + |\downarrow \rangle)$). From here on, we measure the quantum observable $\langle \sigma^z_n \rangle$ for each site $n$, resulting in $\langle \sigma^z_n \rangle = -1$ for a spin down, $\langle \sigma^z_n \rangle = 1$ for a spin up and in between those in case of a local superposition. In the case of the latter, we project the spin to the classical one with the highest probability by using the $\operatorname{sign}$. The resulting spin configuration is further mapped back to the binary encoded solution for each voxel as described in the previous section.

In the case of the IMRT optimization problem, we are dealing with a non-trivial quantum Spin-Glass Hamiltonian type with over 32000 long-range interactions. Thus, the optimization is a highly non-trivial task and for the TTN-algorithm can be quite sensitive to the initialization procedure. Therefore, for each run, we randomly initialize several samples of the TTN from which we start the optimization. In the end, we can verify the best simulation by comparing the resulting energies.

\section{Analysis}
\label{sec:results}

In this section, we compare the cancer treatment optimization performed with the TTN approach against QP and SA. In particular, we show the applications for two different toy models to validate our approach followed by a more realistic anatomical scenario simulating a prostate IMRT treatment. 

\subsection{Methods}
We first compare the TTN algorithm with the analytical solution for a 3D box toy-model scenario. For the second, the two-sphere model and the more realistic IMRT phantom, we lack a general analytical solution. Therefore, we evaluate the results of our TTN approach by comparing it with two different optimization methods, QP and SA. In particular, QP  refers to a set of methods and algorithms used to solve quadratic optimization problems subject to linear constraints and it was exploited to address the initial optimization problem in Eq.~(\ref{eq:cost_function}). On the other hand, the problem in the Ising-type formulation in Eq. (\ref{eq:ising_type}) is addressed using both TTN and SA. QP and SA are used to validate the results obtained with the TTN approach. For further details on the QP and SA, we refer to App.~\ref{app:classical-optimization}.

We point out, that both SA and TNs algorithms contain elements of randomness: samples of independent and randomly initialized simulations are always collected and the best solution considered. This also allows us to calculate the standard deviation of the samples and have a quantitative idea of the general behaviour of the algorithms.

\subsection{Toy Models}

\paragraph*{Bipartite box.} In this section we describe a simplified analytical model used to validate the correctness of the mapping discussed in Sec.~\ref{subsec:mapping} and further our TN approach.

\begin{figure}
    \centering
    \includegraphics[width = \linewidth]{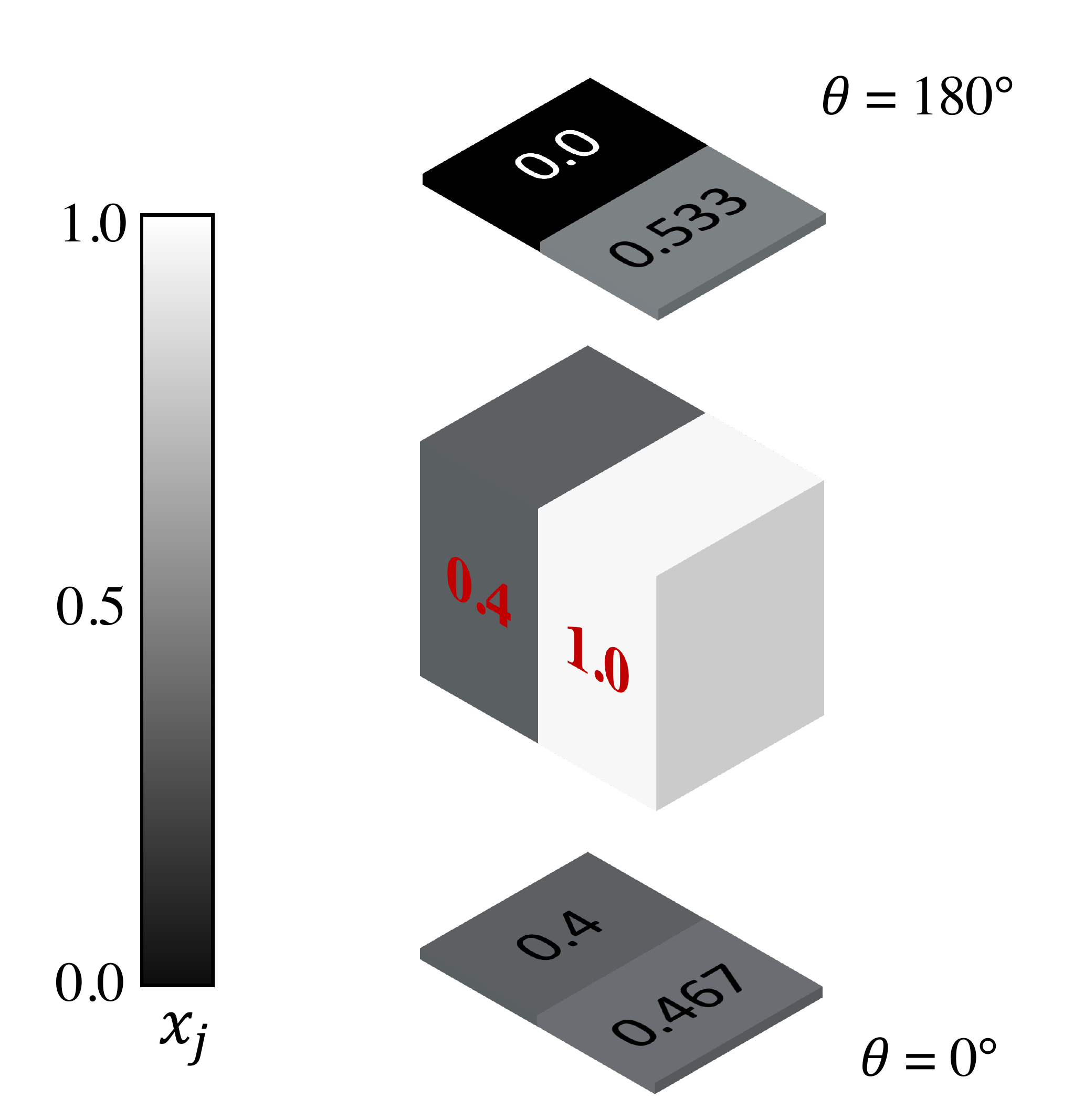}
    \caption{\textbf{Optimization on the 3D box toy model.} The box is radiated from two opposite angles $\theta = 0^{\circ}$ and $\theta = 180 ^{\circ}$. The red numbers on the box are the desired beamlet weights; the number on the upper and lower rectangules are the beamlet weights obtained with TTN. Their sum for each partition equals the desired values.}
    \label{fig:3D-box}
\end{figure}

The model consists of a 3D box subdivided into two different regions for which we assign specific dose prescriptions as shown in Fig.~\ref{fig:3D-box}. The red number on the front of the box corresponds to the desired beamlet weights for each bipartition. We radiate the cube with two beams from two opposite directions  (i.e. $\theta_1=0^{\circ}$, here represented as two rectangles, with a variable number of beamlets $N_B$ for each beam, two in this example. The radiation beam is modelled as an \textit{ideal beam} which releases the same amount of dose to each voxel. Moreover, we neglect scattering effects limiting the interactions among different beamlets to those acting geometrically on the same voxels only. The dose prescriptions and the number of voxels are chosen in a way to ensure the exact ground state with energy $E_{0} = 0$. Therefore, for instance, we choose dose values which are compatible with the number of discretization levels $2^{N_Q}$ used in Eq.~(\ref{eq:decimal_binary}).

We point out, that these conditions are not fulfilled in general cases. Anyhow, we introduce them to obtain an exactly solvable model for the sake of validation. Later on, we present a more realistic anatomical scenario of cancer treatment. The model is then described by the influence matrices $A_r$ for each region $r$, the priorities $\gamma_i$ and the dose prescriptions $D_i^{(P)}$ for each voxel. These are the same information which can be extracted from a real therapy planning system, as we will see also in the following.

The number of bits used to represent each beamlet is set to $N_Q = 4$. The target dose prescriptions for the two partitions are $D_{left}^{(P)} = 6.0$ and $D_{right}^{(P)} = 15.0$, which are arbitrary number chosen according to the number of discretization levels for the $x_j$, 16 in this case.  The influence matrix, mapping the beamlets intensities to the voxels, is defined to have a uniform dose release into the box. Thereby, as we simplified physical effects of the beam, we end up with the influence matrix consisting of either zero and otherwise constant entries mapping the beamlet intensities to the voxels.

The values contained inside the upper and lower and rectangles correspond to the beamlet weights obtained with the TTN algorithm. Wo observe that their sum equals the desired values for each bipartition, 
showing a perfect agreement between the analytical and numerical solution. This proves the well functioning of the algorithm as well as the correctness of the mapping procedure.

We point out, that, in general, there can be more than one configuration satisfying the constraints: indeed, depending on the system parameters, the ground state of the Ising-Hamiltonian can be degenerate. Anyhow, for the treatment, we are satisfied obtaining anyone of the degenerate ground states.

\begin{figure}
\centering
\begin{minipage}{.35\linewidth}
  \centering
  \includegraphics[width=\linewidth]{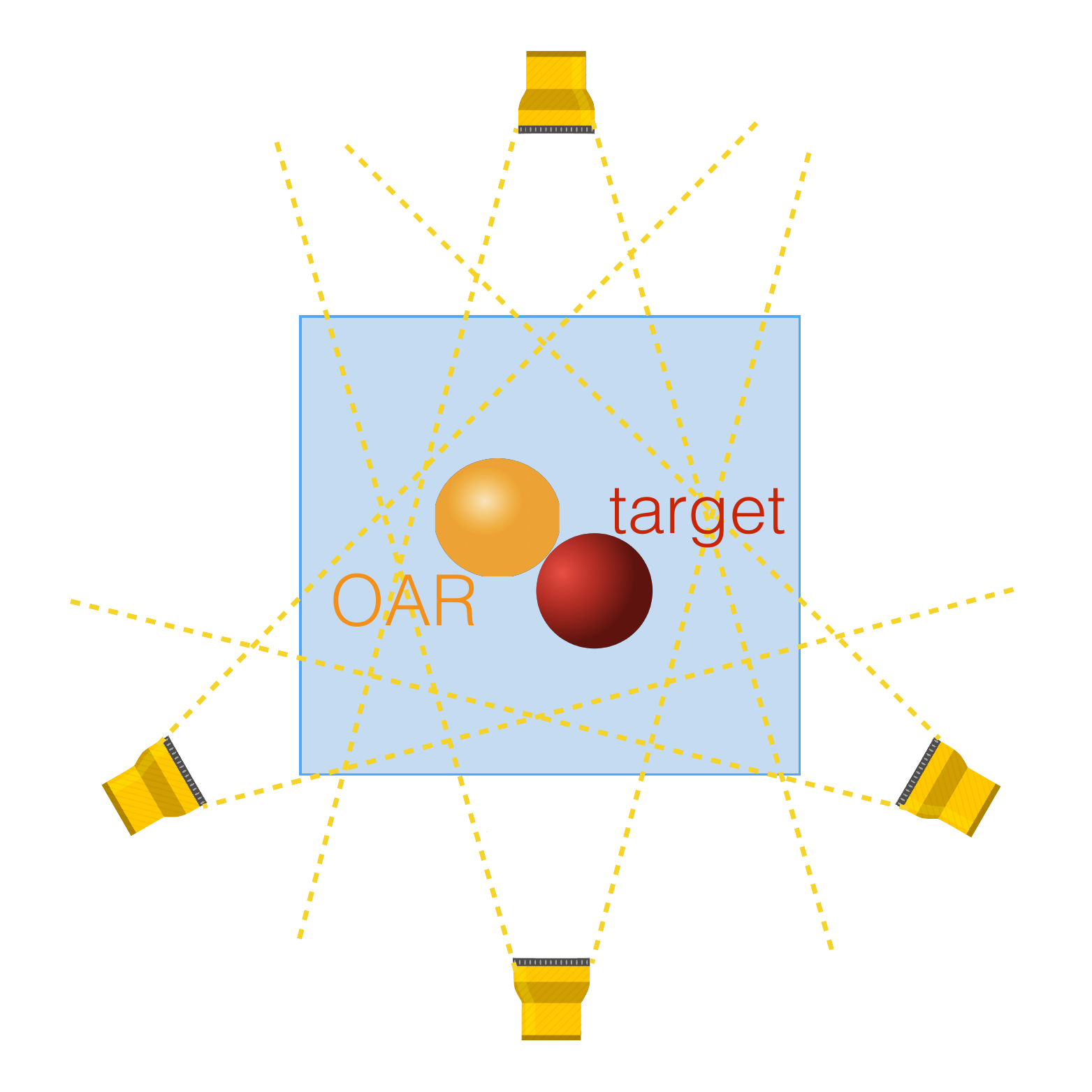}
  \subcaption{}
  \label{fig:optim_sphere_TTN_a}
\end{minipage}%
\hspace{0.02\linewidth}
\begin{minipage}{.58\linewidth}
  \centering
  \includegraphics[width=\linewidth]{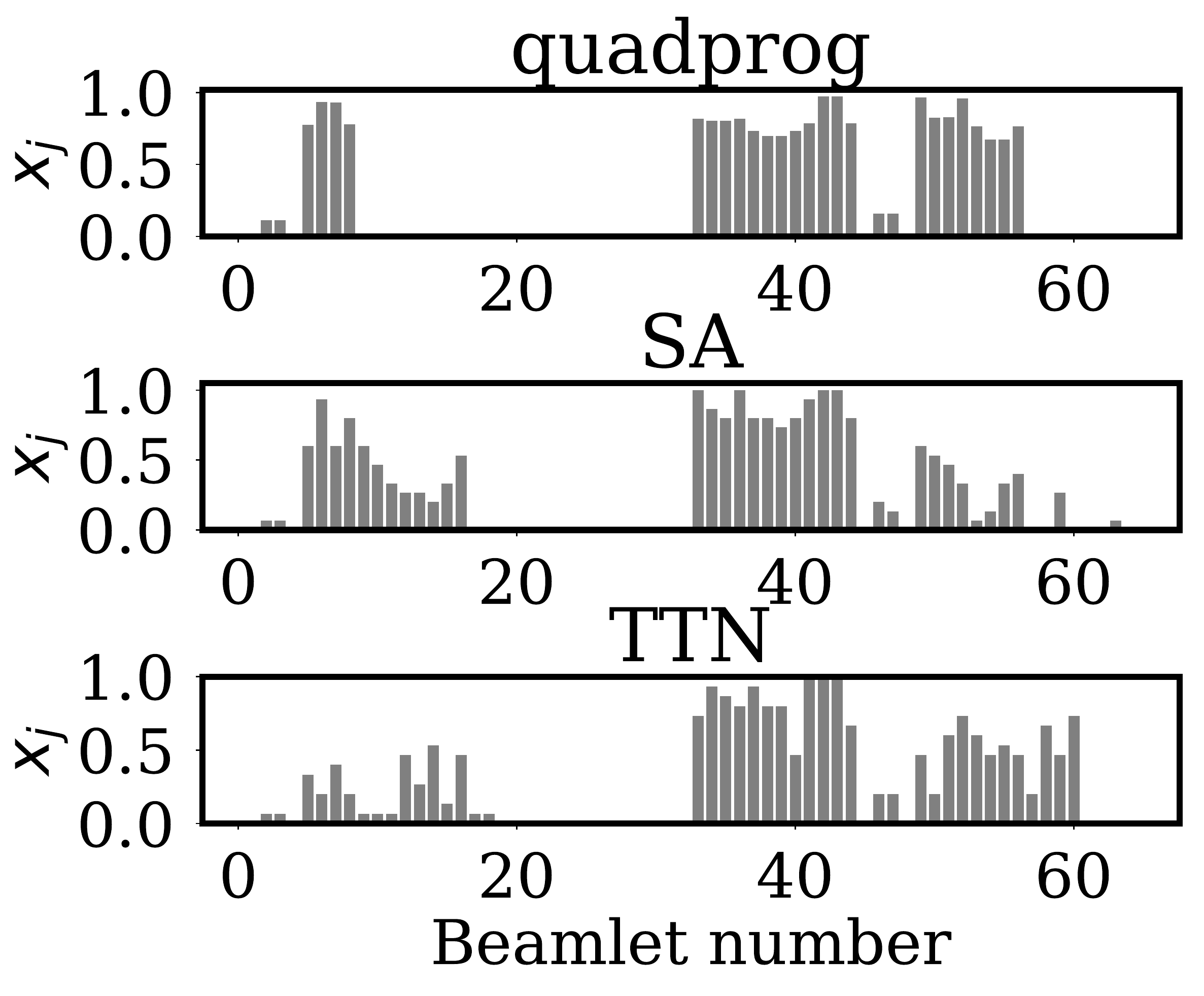}
  \subcaption{}
  \label{fig:optim_sphere_TTN_b}
\end{minipage}
\begin{minipage}{1.0\linewidth}
  \centering
  \includegraphics[width=1.00\linewidth]{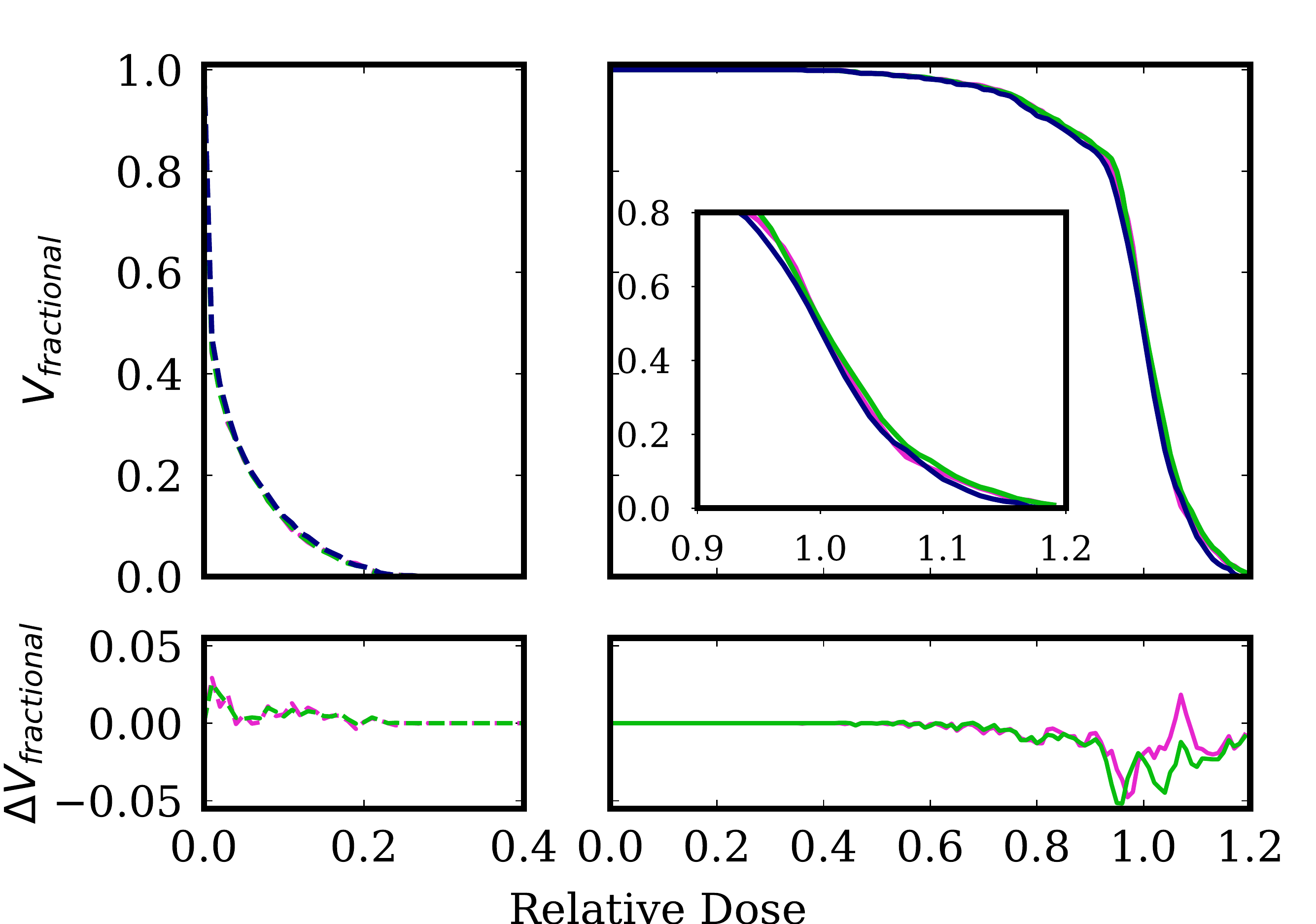}
  \subcaption{}
  \label{fig:optim_sphere_TTN_c}
\end{minipage}%
\caption{\textbf{Optimization on the sphere Toy Model.} Schematic representation of the model with the the red sphere as target and the orange sphere as OAR;(b) best final beamlet configuration for the three algorithms; (c, top) Cumulative DVHs obtained with the three algorithms: quadprog (pink), SA (green) and TTN (blue). On the left panel the orange sphere (dashed line); on the right panel the red sphere (solid line). (c, bottom) Difference in the relative volume coverage between TTN and quadprog (pink line), and TTN and SA (green line).}
\label{fig:optim_sphere_TTN}
\end{figure}

\paragraph*{Sphere.}\label{par:sphere} In this paragraph, we illustrate the application of the TTN algorithm to a more realistic clinical scenario. In this case, instead of assuming an \textit{ideal beam} we model the photon beam using the Matlab-based software CERR \cite{CERR}. This software generates radiotherapy plans and can be used to obtain the influence matrixes $a_{ij}$. It allows working with physical effects introduced by the beams such as scattering and further allows for higher freedom in choosing the geometry, the number of beams and other typical model parameters. Thereby, the underlying medical images are in the standardised DICOM format. 

In Fig.~\ref{fig:optim_sphere_TTN_a} we illustrate the model analysed in this paragraph. The model described consists of a cubic box dimensions $(50.0\times50.0\times100.0)$ cm$^3$ filled with water in which two spherical regions, shown in red and in orange respectively, of diameter $d = 3$ cm placed, one considered as the targeted tumor and the other one as an organ at risk. We irradiate the box using four beams at 0$^{\circ}$, 120$^{\circ}$, 180$^{\circ}$and 240$^{\circ}$ and a total amount of $N_B = 64$ beamlets (16 beamlets/beam). The influence matrices ($A_{red}$ and $A_{orange}$) are obtained using the CERR's dose calculation algorithm QIB \cite{QIB} with the default settings; the dimensions of the beamlets are set $(1.0\times 1.0)$ cm$^2$.

The optimization goals are set to $D_{red}^{(P)} = 50.0$ Gy for the red sphere and $D_{orange}^{(P)}= 0.0$ Gy for the orange one, considering the first as the targeted tumor and the second as an OAR. Each sphere is weighted equally in the cost function with $\gamma = 1.0$. 
During the optimization procedure, the influence matrixes are always normalized to keep the final beamlets intensities in the interval $ \left[0,1\right]$.

In the mapping to the discrete problem, the bit-depth was fixed to $N_Q = 4$ bits, resulting in a fully-connected lattice of $256$ sites for the underlying Hamiltonian. The total number of non-zero interaction terms was $n_{int} = 32640$. Thus, the underlying quantum many-body system is a challenging long-range spin-glass Ising model to be solved with the TTN algorithm.

In practice, it is unfeasible to obtain $E_{0}=0$ and thereby to reach exactly the prescribed dose distribution. Thus, in this example, the optimization balances the different goals for each organ according to their priorities $\gamma$. For this reason, to evaluate the quality of the results returned by the TTN algorithm, the same optimization task was attacked using the Matlab build-in function \textit{quadprog}, which exploits QP to optimize the cost function, and SA. We recall that the optimization with \textit{quadprog} was directly performed on the function in Eq.~(\ref{eq:cost_function}), while SA and the TTN were applied to the discretized problem in Eq.~(\ref{eq:ising_type}).

A standard method used for plans quality evaluation is the cumulative DVH histogram, which shows a 2D projection of the 3D dose distribution inside a given volume. It represents the fractional volume receiving at least a given value of dose. Given a generic volume, $r$, we can easily build the dose vector {\bf D$_r$(x)} as described in Eq.~(\ref{eq:delivered_dose}) by applying its influence matrix, $A_r$, to the beamlets vector {\bf x}. The resulting vector contains the total dose delivered to each one of the voxels in the volume r. By subdividing the dose interval $[0,D_i^{(max)}]$ into $n_b$ (dose) bins, for each of them we can count how many voxels receive a dose greater or equal than the corresponding dose value. In other words, the number of entries in the k-th bin indicates the number of voxels receiving at least the corresponding dose. The obtained distribution results in a cumulative DVH with the fractional volume represented on the y-axis and the dose values on the x-axis. We point out, that in this representation of the dose distribution we lose the spatial information of the problem. 

On the top panels of Figure~\ref{fig:optim_sphere_TTN_c}, we show the DVHs obtained with the three methods. It is clearly visible, that the three methods show a very good qualitative agreement in the resulting DVHs for each organ. This agreement is further quantitatively confirmed by the energy $E_{0}$ - or cost - after the \english{minimization}: Within the statistical uncertainty, all three methods result in a final energy $E_{0} = 0.0181(8)$. By looking at the bottom panels of Figure~\ref{fig:optim_sphere_TTN_c} we see that the difference between TTN and the other two methods in the relative volume coverage is globally very close to zero, with only a few peaks at about 2-5 $\%$

In Figure \ref{fig:optim_sphere_TTN_b}, we present the final beamlets configuration for each of the three methods. Despite their global consistence, we observe that local differences arise. First, this is due to the fact the final configurations have slightly different energies, despite are all consistent. However, in general, there may exist more than one configuration satisfying the constraints and minimizing the energy $E_{0}$ for the underlying system and this effect is further amplified when comparing the optimization on the discrete space (SA and TTN) to that on the continuous one (quadrog). This happens because the energy landscape may be altered by the dicretization procedure. Thus, in this case, we have that the ground state of the underlying Hamiltonian is either degenerate or its energy gap is reasonably small. In fact, the underlying quantum spin-glass Hamiltonian has a highly non-trivial spectrum with many local optima and depending on the system parameters degenerate ground-states.

\subsection{Prostate cancer treatment with TG119 IMRT phantom}
\begin{figure}
\centering
\begin{minipage}{0.4\linewidth}
  \centering
  \includegraphics[width=\linewidth]{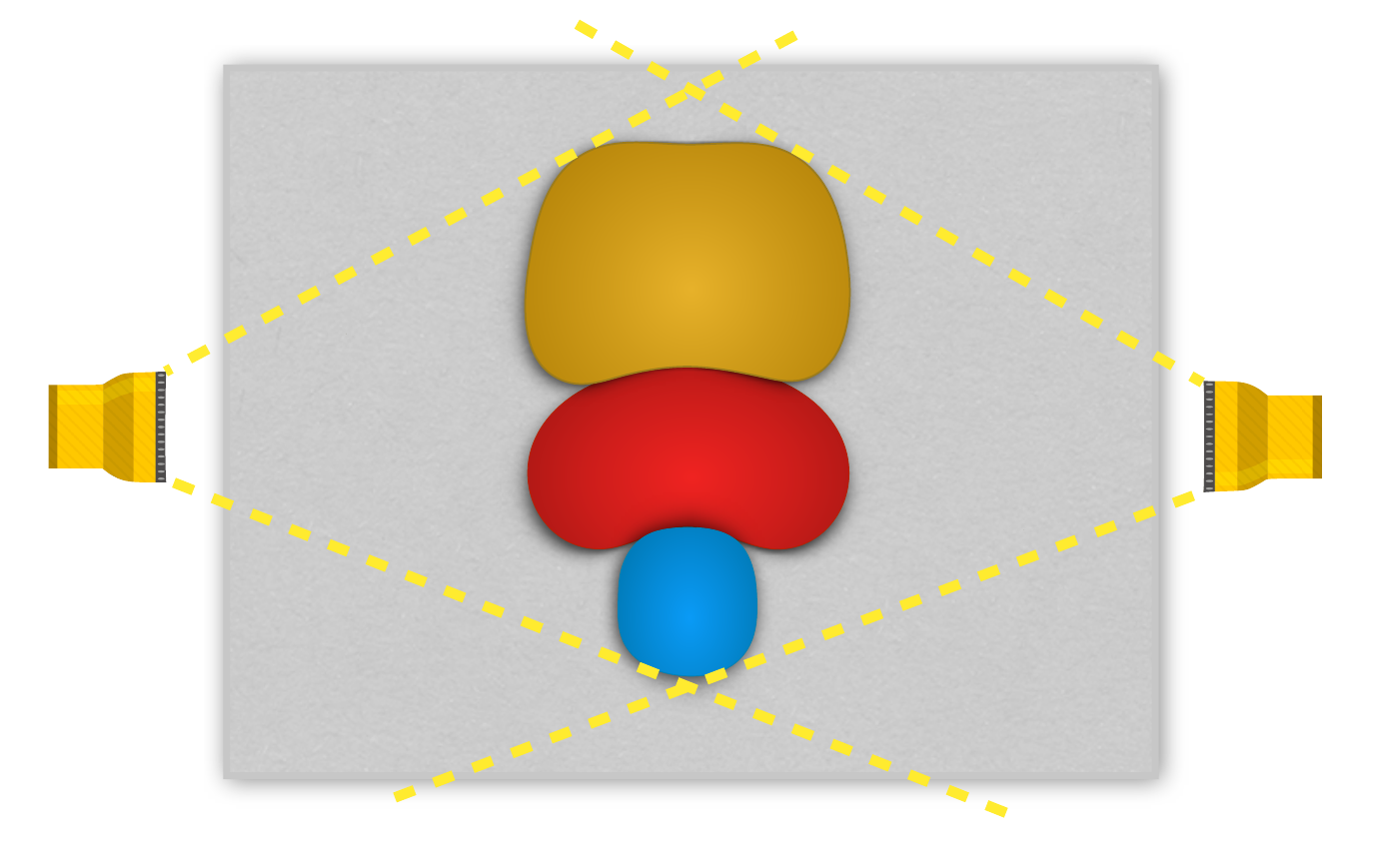}
  \subcaption{}
  \label{fig:prostate-model}
\end{minipage} 
\begin{minipage}{0.58\linewidth}
\centering
  \includegraphics[width=\linewidth]{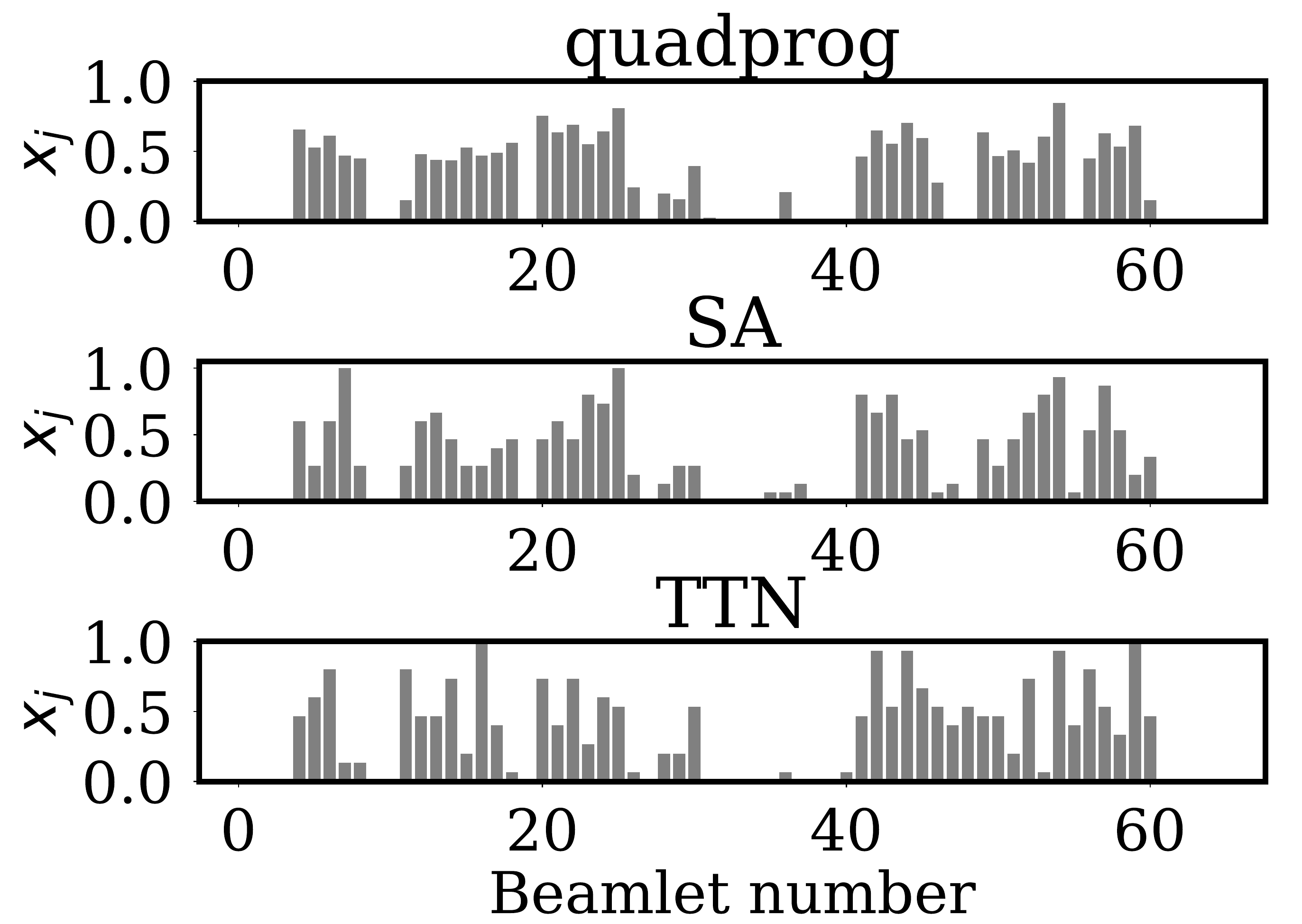}
  \subcaption{}
  \label{fig:optim_prostate_beamlets}
\end{minipage}

\begin{minipage}{1.0\linewidth}
\centering
\includegraphics[width=1.02\linewidth]{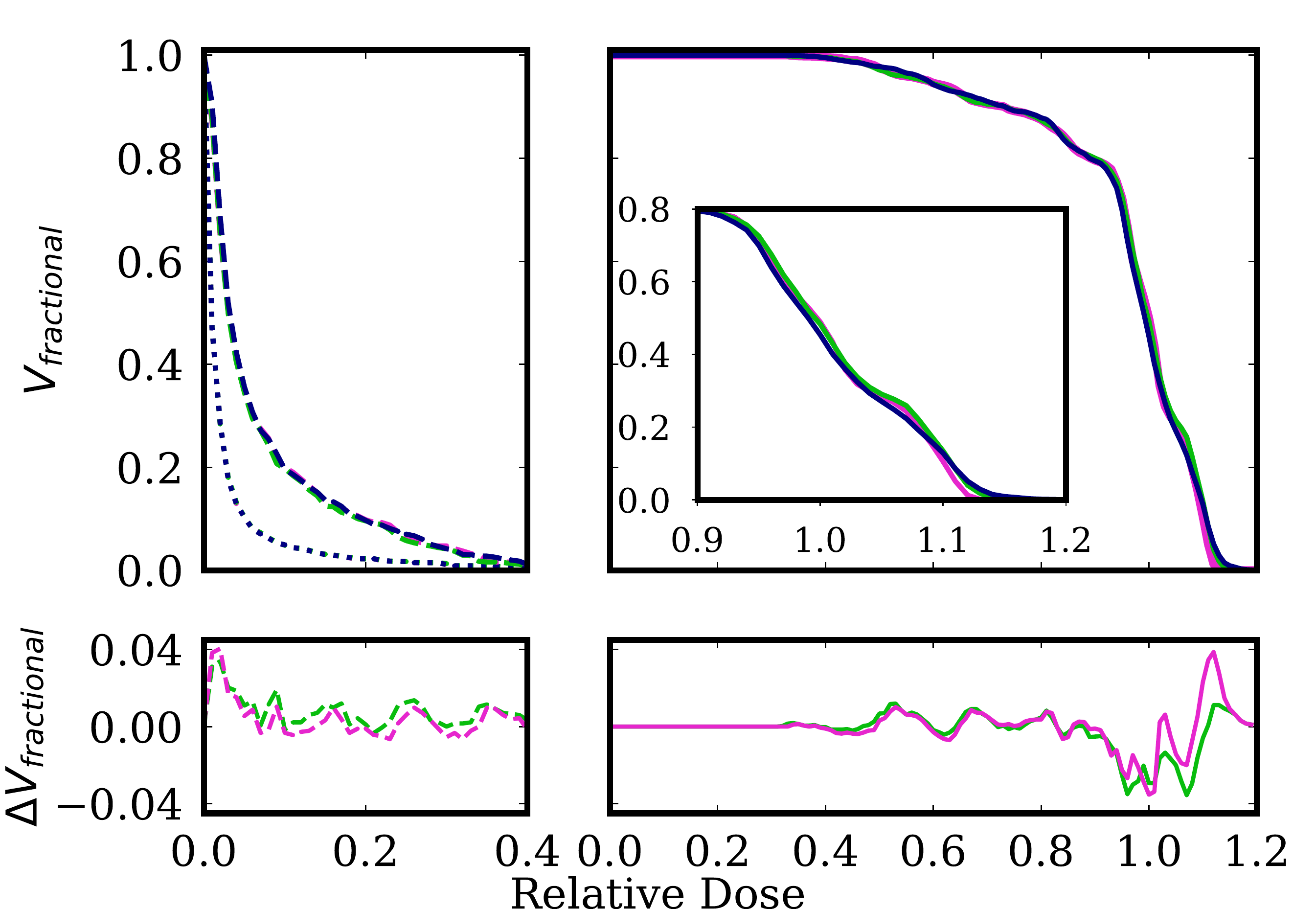}
  \subcaption{}
  \label{fig:optim_prostate_DVHs}
\end{minipage}
\caption{\textbf{Optimization on the TG119 IMRT phantom.}(a) Schematic representation of the model with the the prostate as target (red) and the bladder (brown) and the rectum (blue) as OARs;(b) Best final beamlets configuration for the three algorithms; (c, top) Cumulative DVHs obtained with the three algorithms: quadprog (pink), SA (green) and TTN (blue). On the left panel the two OARs: bladder (dotted line) and rectum (dashed line). On the right panel the prostate (solid line). (c, bottom) Difference in the relative volume coverage for the rectum (dashed line) and the prostate (solid line) between TTN and quadprog (pink line), and TTN and SA (green line).}
\end{figure}

We now show the results obtained on a standard IMRT phantom provided by the American Association of Physicists in Medicine Task Group 119 for use in institutional IMRT commissioning~\cite{TGdataset, DatabaseProstate}. This dataset contains several segmented structures and we chose the following with the aim of simulating a prostate cancer cases: prostate as the targeted tumor, bladder and rectum as the OARs.

The geometry we used is characterized by two beams (33 and 31 beamlets) placed at 90$^{\circ}$, 270$^{\circ}$ as shown in Fig. \ref{fig:prostate-model}, resulting in a total amount of $N_B = 64$ beamlets. The dimensions of each beamlet were $\left(0.9\times0.9\right)$ cm$^2$. The dose prescriptions were set to $D_{prostate}^{(P)} = 50$ Gy and $D_{bladder}^{(P)} = D_{rectum}^{(P)} = 0.0$ Gy and the priority assigned to the different structures was $\gamma = 1.0$ for all of them. 
The dose calculation algorithm used was CERR's QIB algorithm in the default settings. The fraction of non-zero elements in the influence matrixes was $0.55$ for the $A_{prostate}$, $0.28$ for $A_{bladder}$ and $0.45$ for $A_{rectum}$.
For the discrete problem, the system obtained is again characterized by 256 fully-connected long-range interacting spins and a total amount of $n_{int} = 32640$ interaction terms. 

The top panels of Fig. \ref{fig:optim_prostate_DVHs} shows the comparison between the DVHs for the three optimization methods. The bottom panels show that the difference between TTN and the other two methods in the volume coverage for the rectum and the prostate is within the 4$\%$, proving a very good quantitative agreement between them. The results for the bladder are not shown since the differences were negligible.  This agreement is additionally confirmed by the obtained energy $E_{0}$ - or cost - after the \english{minimization}: within the statistical uncertainty, the three methods result in a final energy $E_{0} = 0.043(1)$. This result further confirms what found from the study on the toy models.

\section{Conclusions}\label{sec:conclusion}

In this manuscript, we presented a new  approach based on TNs to optimize the dose distribution for an IMRT cancer treatment. We showed a feasibility study on three different cancer treatment scenarios. First, we provided a proof-of-principle 
Tensor Network analysis by successfully investigating an analytically solvable toy-model of a radiated box. Then, we compared the TNs with the classical approaches of QP and SA in the case of a spherical cancer and a spherical organ at risk.
Finally, we illustrated the successful application of the TNs approach to a more realistic anatomical scenario simulating a prostate cancer.

The main goal of our work was to show the applicability of TNs to the IMRT optimization problem, fostering new applications of quantum-inspired techniques to the solution of classical optimization problems. Along the road, we defined a clear strategy to map the classical problem to simulated quantum-like hardware. 

Our results indicate that the TN approach can achieve results compatible with other optimization techniques such as QP and SA. We stress that for this feasibility study we (i) used a reduced number of beamlets to reduce the complexity in the models, (ii) kept the cost function convex for sake of simplicity and (iii) used the TNs code "out-of-the-box" originally engineered for typical quantum systems with significantly fewer interactions but higher entanglement. Further software developments will allow to address these three points, increasing the TN approach efficiency. In particular, extending this study to non-convex and non-differentiable functions and by further specializing and parallelizing the TNs code for this particular field of application, further significant steps forward can be made towards real-world scenarios and their use in every-day medical care. 

Finally, we point out that TNs are particular examples of quantum circuits, thus, this study opens the way to the application of quantum computation to cancer treatment, for example through the application of hybrid quantum-classical optimization algorithms~\cite{McClean_2016}. Once quantum computer hardware will be scaled up, one could replace the TNs simulation with actual quantum computation, possibly further enhancing our capabilities of fighting cancer via IMRT.

\section{Acknowledgments}

This work is partially supported by the Italian PRIN 2017 and Fondazione CARIPARO, the Horizon 2020 research and innovation programme under grant agreement No 817482 (Quantum Flagship - PASQuanS), the QuantERA projects QTFLAG and QuantHEP, and the DFG project TWITTER. We acknowledge computational resources by CINECA, the Cloud Veneto, the BwUniCluster and by the ATOS Bull HPC-Machine.

\appendix

\section{Classical optimization}\label{app:classical-optimization}

\subsection{Quadratic programming}
Quadratic programming (QP) refers to a set of widespread methods for solving (non-linear) quadratic optimization problems subject to linear constraints. A general QP problem can be formulated as follows:
\be \label{eq:QP}
\min_x\  {\bf q}^T {\bf x} + \frac{1}{2}{\bf x}^T Q {\bf x} \ \  \ \ s.\ t. \ \ \ \ \begin{cases}
A{\bf x} = {\bf a}, \\
B{\bf x} \leq {\bf b} \\ 
{\bf x}\geq {\bf 0} 
\end{cases}
\ee
where $f({\bf x}) $ is the target function, {\bf q} its gradient and $Q$ its Hessian matrix. The equation $A{\bf x} = {\bf a}$ contains all the equality constraints, while $B{\bf x} \leq {\bf b}$ all the inequality constraints. 

In the QP formulation, the initial IMRT optimization problem in Eq.~(\ref{eq:cost_function}) can be written as follows:
\be \label{eq:QP-IMRT}
\min_{\bf x} \ 2\sum_r\left(  {\bf x}^T \tilde{A}_r^T\tilde{A}_r{\bf x} - {\bf D}^{(P)}_r  \tilde{A}_r {\bf x} \right), \  {\bf 0} \leq {\bf x} \leq {\bf 1} 
\ee
where $r$ is an index running over the different volumes or organs (the targeted tumor and the OARs), $\tilde{A}_r$ is the influence matrix for the r-th volume whose entries are weighted by the pre-assigned priorities $\gamma_{i(r)}$ to each voxel, $i$, and ${\bf D}^{(P)}_r$ a vector containing the dose prescriptions for the voxels in the r-th volume.

Matlab's \textit{Optimization Toolbox}\textsuperscript{\texttrademark} provides the function \textit{quadprog} which exploits different solvers to attack a wide class QP problems. In this work, this methods is used to solve the initial quadratic problem and produce results to be compared to those obtained with TTN.  

\subsection{Simulated annealing}\label{app:simulated-annealing}

Simulated annealing (SA) is a widespread combinatorial optimization method based on randomization techniques~\cite{simulated-annealing}, in particular based on the Metropolis-Hastings algorithm. By varying a control parameter T, called \textit{temperature}, it is possible to explore the landscape of a target free-energy function in order to find its global minima. SA is typically applied to large-scale non-convex optimization problems where the number of local minima in the energy landscape is very high. 
The algorithm always requires a starting point which in practice is either a random one or the best one known for a specific problem. From the initial starting point, a rule to generate new configurations is given: Configurations with an energy lower than the previous configuration are always accepted (with probability one). On the other hand, a move towards configurations with a higher energy is accepted with a certain probability which significantly reduces the risk of getting stuck into a local minimum. In this case, the acceptance probability decreases dynamically with the temperature T, which itself  decreases during the optimization from a given value $T_{max}$ to $T_{min}$ according to a pre-defined annealing schedule.

In this work, SA is applied to find the ground state of the classical Ising spin-glass problem representing the initial IMRT optimization problem. We always start the SA from a random configuration for the spins in the lattice. New configurations are generated by flipping a randomly chosen spin in the prior configuration. Due to the intrinsic stochastic nature of SA, we perform a statistical sampling of $N_{runs}$ independent and randomly-initialized optimization runs for the same problem. For both, the sphere and the prostate cancer case, $N_{runs}$ is set to 100.

The code used in this study to perform SA is based on the Python library in~\cite{SA-algorithm}.

\section{Ground-state search via TTN}\label{app:ground-state-TTN}

\subsection{Trend of the solutions}
\begin{figure}[h!]
\centering
\begin{minipage}{1.0\linewidth}
    \includegraphics[width=1.02\linewidth]{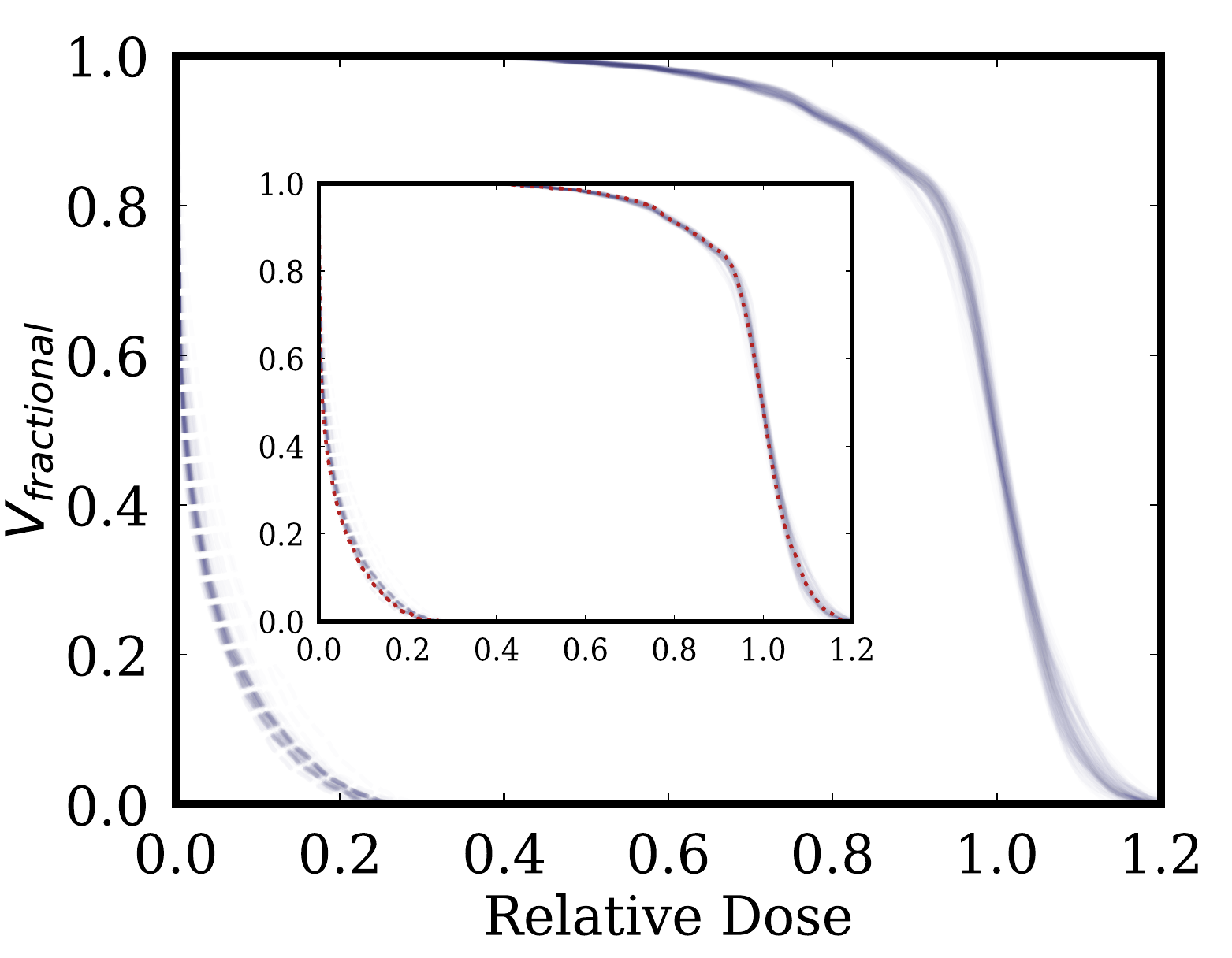}
  \subcaption{}
  \label{fig:TTNvsOTHERS-sphere}
\end{minipage} 
\begin{minipage}{1.0\linewidth}
  \includegraphics[width=1.0\linewidth]{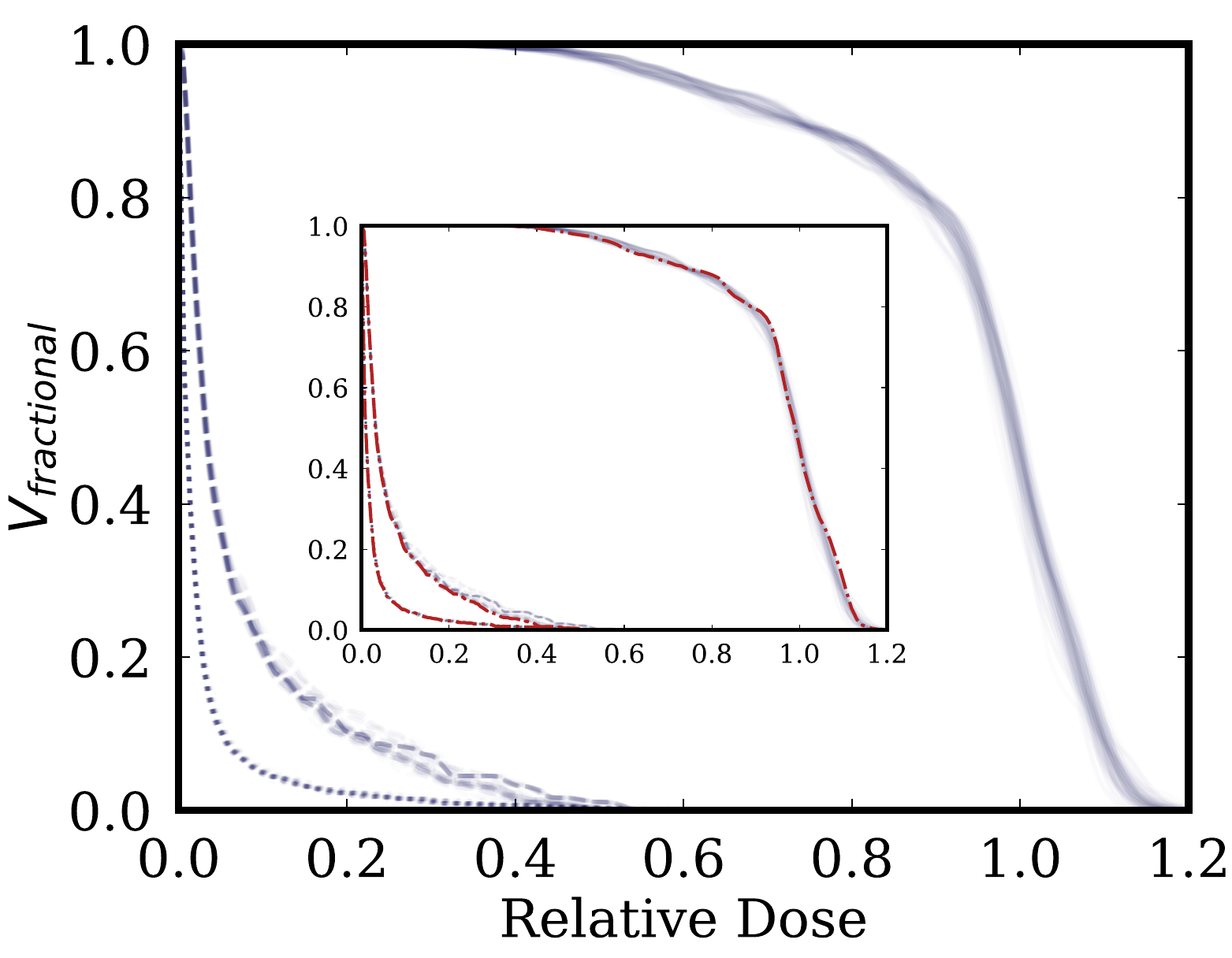}
    \subcaption{}
    \label{fig:TTNvsOTHERS-prostate}
\end{minipage}
\begin{minipage}{1.0\linewidth}
\raggedleft
\includegraphics[width=0.88\linewidth]{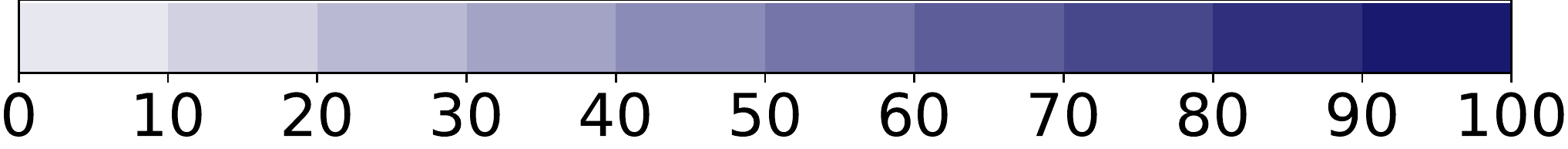}
  \subcaption*{}
\end{minipage}
\caption{\textbf{Trend of the 100 solutions obtained with the TTN algorithm.} (a) Sphere toy-model. (b) Prostate cancer case. The intensity of the blue lines is proportional to the number of superimposed solutions, according to the colorbar. The innermost panels show the comparison between the 100 solutions and the best one.}
\end{figure}
The TTN algorithm explores a corner of the full many-body Hilbert space ${\cal{H}}$ which becomes wider as the bond dimension $\chi$ increases. In this work, the bond dimension was fixed to $\chi = 5$ throughout the whole study. Since we are dealing a classical problem we know that the final solution is not entangled. However, the introduction of a bond dimension $X > 1$ increases the probability for the algorithm to converge to the global minimum, as it thereby explores a greater solution space. 

In Sec. \ref{sec:results} we've shown the best results obtained with the TTN both for the prostate and the sphere over samples of $N_{runs} = 100$ runs. We recall that the tensors entries are randomly initialized at the beginning of each new optimization run.

In order to see which is the general behaviour of the algorithm throughout the whole sampling, we compare in Fig. \ref{fig:TTNvsOTHERS-sphere} and \ref{fig:TTNvsOTHERS-prostate} the results over all $N_{runs} = 100$ runs. We observe that they are globally very closely distributed around the best solution and this proves the precision of the TTN algorithm. However, making precise clinical considerations about the consistency between the different solutions goes far beyond the purpose of this feasibility study since many different factors should be considered depending on the specific case.

\subsection{Reduction of interaction coefficients.}
\begin{figure}[h!]
\centering
\begin{minipage}{1.0\linewidth}
    \centering
    \includegraphics[width=\linewidth]{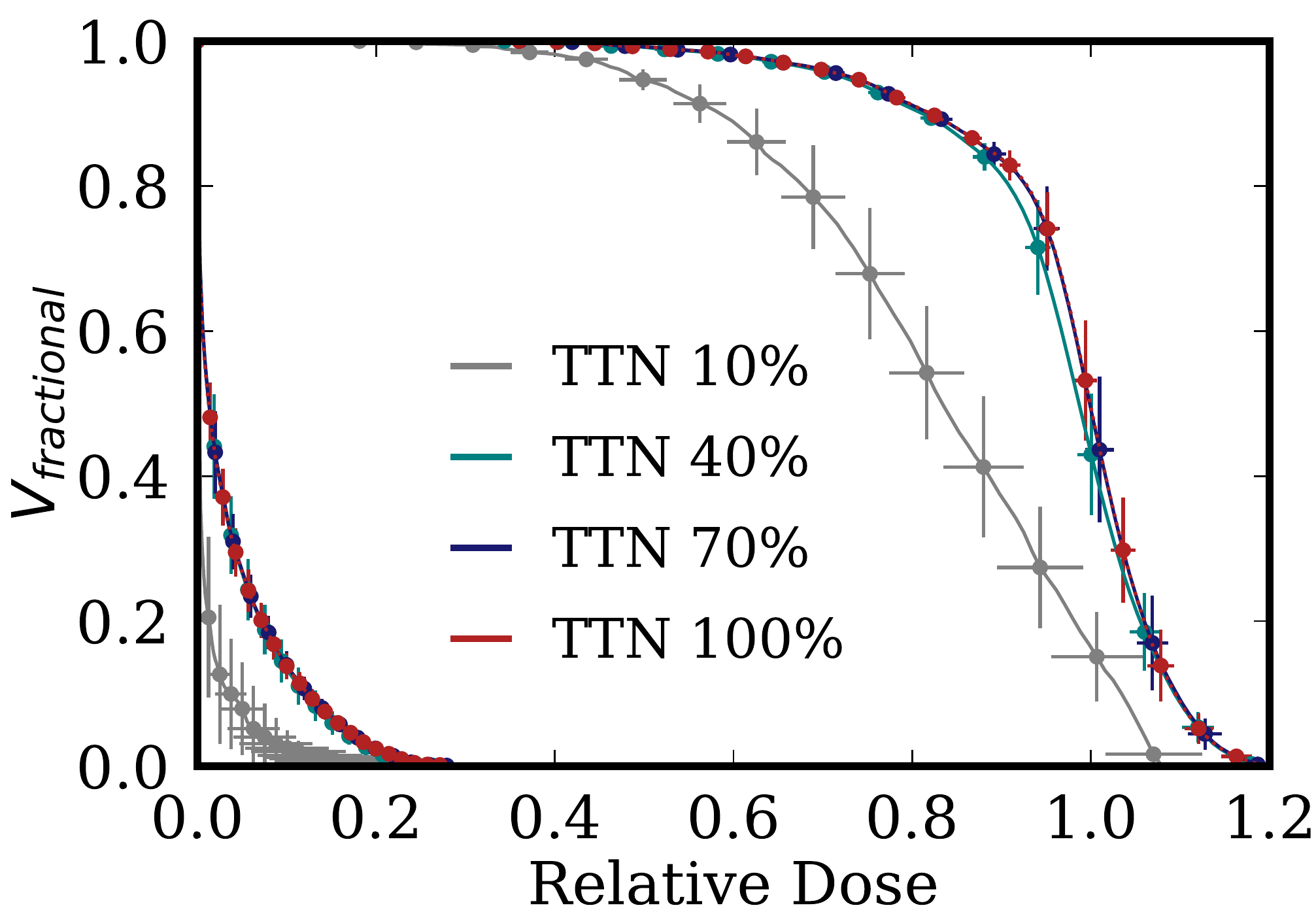}
    \subcaption[]{}
    \label{fig:TTN-compression-sphere}
\end{minipage}
\begin{minipage}{1.0\linewidth}
    \centering
    \includegraphics[width=\linewidth]{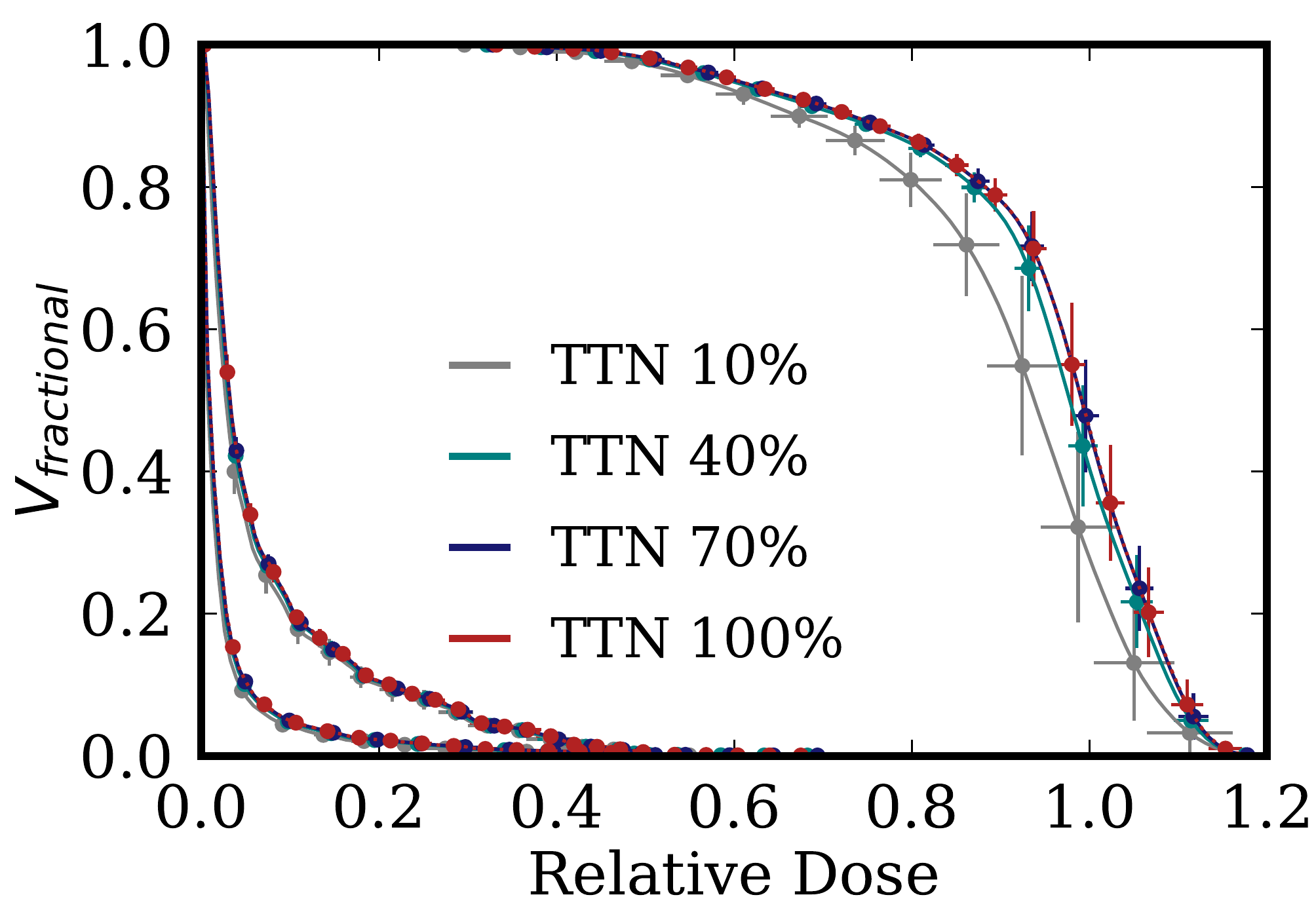}
    \subcaption[]{}
    \label{fig:TTN-compression-prostate}
\end{minipage}
\caption{\textbf{Reduction of the number of interaction coefficients.} (a) Sphere toy-model. (b) Prostate cancer. }
\end{figure}

\begin{figure}[h!]
    \centering
    \includegraphics[width = \linewidth]{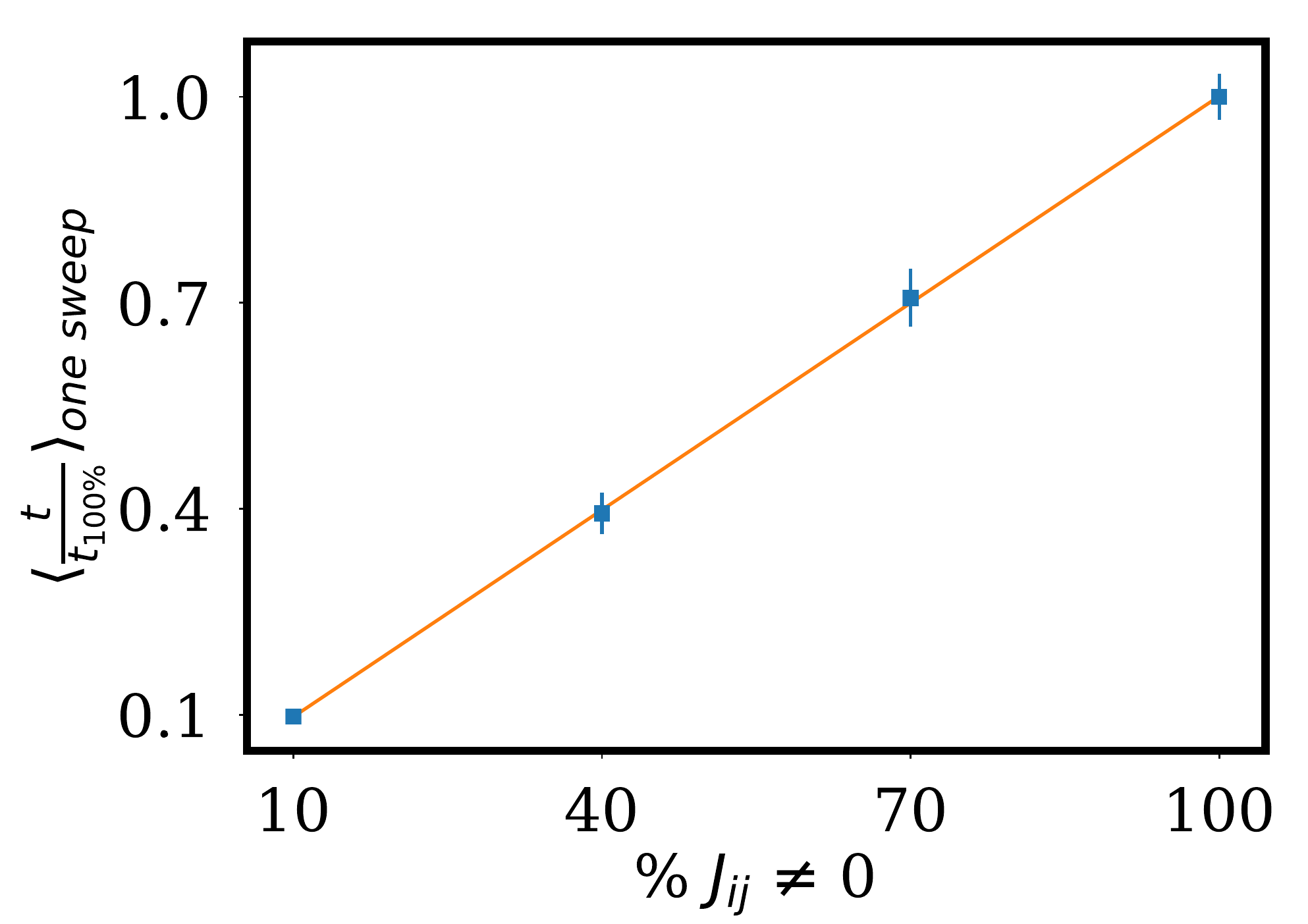}
    \caption{\textbf{Scaling of the computational time.} The figures shows the average computational time to perform one sweep when different fractions of non-zero coefficients are considered. Each point is shown with the associated standard deviation. Each sample was made up of about 700 sweeps. Each time point is normalized by the average time to perform one sweep when the $100\%$ of coefficients is considered. The orange line is the results of a linear fit.}
    \label{fig:time-scalle-prostate}
\end{figure}

The model we have considered so far is a fully-connected lattice system of $N$ interacting spins, with the number of interactions terms given by $N\times(N-1)/2$. The density of the interaction scheme has an impact on the computational time required to the TTN to converge. Thereby, it is interesting to investigate the behaviour of the accuracy of the solution when neglecting some of the coefficients. In particular, we fix a threshold, $\lambda_{min}$ such that $(|J_{ij}| < \lambda_{min}) = 0$, with $J_{ij}$ the pairwise interaction term between the i-th and the j-th spin in the lattice, to cut away the interaction terms with lowest coefficients.  

In the following, we considered four different cuts, keeping respectively the $10\%$, $40\%$, $70\%$ and $100\%$ of the coefficients. The result for the sphere cancer is shown in Fig. \ref{fig:TTN-compression-sphere}. For each of the four cases, samples of $N_{runs} = 100$ runs were collected. Each histogram shows the average position of each bin (we set $N_{bin} = 100$ bins) with the associated standard deviation both on the relative dose and the factional volume axes. For the red sphere, (the most-right lines) $70\%$- and $100\%$- results are very well superimposed, while the $40\%$-line lays within the standard deviation of the former ones. Bigger differences arise between these three and the $10\%$ line. No significative differences can be found for the orange sphere (most-left lines) between the $100 \%$, $70\%$ and $40\%$ lines. Even in this case, the main differences arise between the $10\%$ line and the others. 

Interestingly, these results confirm that accurate solutions can be obtained even considering only $40\%$ of the coefficients.  The study on the prostate cancer case (see Fig. \ref{fig:TTN-compression-prostate}) further confirms this insight. However,in this case, the $10 \%$ line is closer to the others than in the previous case. 

This result becomes even more interesting if we consider the scaling of the computational time when reducing the number of interaction coefficients. Fig. \ref{fig:time-scalle-prostate} shows the result for the prostate cancer case. We observe that the scaling is linear. In conclusion, using this procedure we should be able to obtain very accurate solutions spending only the $40\%$ of the initial computational time. 

Consequently, one question arises naturally: How to find the optimal balance between speeding up the simulations by cutting terms versus keeping an adequate information to accurately describe the system? 
We therefore propose the following heuristic procedure:
\begin{itemize} 
\item Build the fully-connected interaction matrix of the initial problem, $J_{ij}^0$ and compute its eigenvalues $s_0$.
\item Choose a cut in order to keep the $\eta \%$ of the coefficients and compute the new eigenvalues $s_{red}$ of the reduced interaction matrix, $J_{ij}^{red}$;
\item Check the statistical difference between the two samples of eigenvalues $s_0$ and $s_{red}$ (before and after the cut). This can be performed using of an hypothesis testing procedure, where the null-hypothesis $H_0$ is that the two sets of eigenvalues are sampled from the same population, hence no statistical differences arise between them. Within this framework, the p-value of the test can  be used to predict the accuracy of the optimization. In particular, the smaller the p-value is, the more the two sets of eigenvalues are likely to come from different populations and thus the accuracy of the optimization on the reduced model to be poor. We point out, that the choice of the statistical test in this procedure is not unique. Our trials exploited the Wilcoxon non-parametric test for dependent samples.
\end{itemize}

\bibliographystyle{apsrev4-1}

\bibliography{references.bib}

\end{document}